\documentclass[letter]{aa}
\usepackage{txfonts}
\usepackage{graphicx}
\usepackage{amssymb}
\usepackage{epsfig}
\usepackage{natbib}
\usepackage{tabularx}
\usepackage{textcomp}
\usepackage{color}

\newcommand{\ebv}{$E_{B-V}$}

\newcommand{\be}{\begin{equation}}
\newcommand{\ee}{\end{equation}}

\newcommand{\kms}{\mbox{km\ \ensuremath{\rm{s}^{-1}}}}
\newcommand{\SN}{\mbox{\object{SN2006X}}}

\begin{document}

\title{Interstellar atoms, molecules and diffuse bands\\ toward SN2006X in M100~\thanks{%
Based on observations with VLT-UVES under Run IDs 276.D-5048, 277.D-5003 and 277.D-5013
and CFHT-ESPaDOnS run 05ao5.}}

\author{Nick L.J. Cox\inst{1} and Ferdinando Patat\inst{2}
}
\authorrunning{N.L.J. Cox \& F. Patat}
\titlerunning{Interstellar medium toward \SN\ in M100}

\offprints{Nick Cox \email{nick.cox@esa.int}}

\institute{Herschel Science Centre, European Space Astronomy Centre, 
ESA, P.O.Box 78, E-28691 Villanueva de la Ca\~nada, Madrid, Spain.
\and
European Southern Observatory, K. Schwarzschild Str.\,2, 85748, Garching b. M\"unchen, Germany
}

\date{Received 17 March 2008 / Accepted 6 May 2008}

\abstract{%
}{%
Supernovae offer the unique possibility to probe diffuse extra-galactic sightlines 
via observation of the optical transitions of atoms, molecules and the 
diffuse interstellar bands (DIBs). Through optical spectroscopy the presence of (complex) molecules
in distant galaxies can be established and used to derive local physical conditions of the 
interstellar medium (ISM).
}{%
High resolution optical (3300 -- 6800~\AA) spectra of \SN\ at different phase obtained with UVES 
on the VLT were reduced and analysed.
}{%
In addition to previously detected atomic (\ion{Na}{i} and \ion{Ca}{ii}) and molecular (CN) transitions
we present detections of DIBs ($\lambda\lambda$6196, 6283), diatomic molecules 
(CH, CH$^+$) and neutral atoms (\ion{Ca}{i}) in the spectra of \SN\ taken at different phases (at 2 days before 
and 14 and 61 days after the brightness maximum). 
An analysis of the absorption profiles shows no variation between phases in the abundance, nor the central
velocities (within 3$\sigma$ error bars) of the (dense) gas tracers (CH, CH$^+$ and \ion{Ca}{i}) and the DIBs.
This is consistent with the conclusion in the literature that \SN\ exploded behind a dense interstellar
cloud (inferred from strong atomic sodium and calcium lines and CN transitions) which caused strong photometric 
reddening but whose material was not directly affected by the supernova explosion.
The CH and CN column densities correspond  to a reddening of one magnitude following the Galactic correlation
derived previously.
The $\lambda\lambda$6196 and 6283 lines detected in the \object{M100} ISM are under-abundant by factor of 2.5 to 3.5
(assuming a visual extinction of $\sim$2 mag) compared to the average Galactic ISM relationship. 
Upper limits for $\lambda\lambda$6379 and 6613 show that these are at least a factor of seven weaker.
Therefore, the Galactic DIB-reddening relation does not seem to hold in \object{M100}, although the lower 
gas-to-dust ratio may further reduce this discrepancy.
}{%
}

\keywords{ISM: lines and bands -- ISM: molecules -- ISM: dust, extinction -- ISM: clouds -- ISM: individual objects: M100}

\maketitle

\section{Introduction}
Optically bright supernovae (SNe) offer the unique possibility to probe the diffuse interstellar medium (ISM)
in galaxies beyond those of the Local Group. 
Despite the serendipitous nature of supernovae events, the study of their high-resolution optical spectra
for interstellar lines has seen a number of successes in recent years.
Optical studies of narrow interstellar absorption lines and diffuse interstellar bands 
superimposed on the continuum and broad line spectra of supernovae started in earnest in the late 1980s.
First, \citet{1987AJ.....94..651R} detected strong sodium lines and tentatively four diffuse interstellar bands (DIBs)
toward \object{SN1986G} in \object{NGC5128}. \citet{1989A&A...215...21D} confirmed the presence
of DIBs toward \object{SN1986G} and reported on the detection of CH and CH$^+$.
\object{SN1987A} provided a unique opportunity to observe interstellar lines and diffuse 
interstellar bands in the LMC (\citealt{1987A&A...177L..17V}, \citealt{1987A&A...182L..59V},
\citealt{1999ApJ...512..636W}).
\citet{1990AJ.....99.1476S} detected sodium, calcium and diffuse interstellar bands toward \object{SN1989M}
in \object{NGC4579}.
Complex sodium and calcium profiles were observed toward \object{SN1993J} in \object{M81} 
but no DIBs or molecules were detected (\citealt{1994A&A...291..425V}, \citealt{1994A&A...285L..13B}).
\citet{2005A&A...429..559S} detected and resolved most of the prominent diffuse interstellar
bands toward \object{SN2001el} in the spiral galaxy \object{NGC 1448}.
Their high-resolution spectra showed a remarkable similarity (both in relative strength and profile) 
between DIBs observed in this distant ($\sim$15~Mpc) galaxy and the Milky Way.
GRBs and DLAs have been used to study the ISM in even more distant systems
(e.g. \citealt{2007MNRAS.tmpL.124E}, \citealt{2007arXiv0711.1372L}).
In contrast, the most distant galaxy probed by direct observations of early-type supergiants is \object{M31}
\citep{2008A&A...480L..13C}. 

The spectra discussed in this paper were previously investigated in the context of the presence and evolution of
circumstellar material of supernovae \citep{2007Sci...317..924P}.
Motivated by the report of \citet{2006CBET..421....1L} on strong interstellar features we
further exploit these spectra to characterise and analyse the ISM
of \object{M100} probed by \SN.
First, we give a brief overview of relevant data on \SN\ (Sect.~\ref{sec:sn}) 
and outline the observations (Sect.~\ref{sec:observations}).
Next, we present the detected (di)atomic lines and DIBs and derive column densities and equivalent widths
(Sect.~\ref{sec:results}).
The results and implications are discussed in Sect.~\ref{sec:discussion}.

\section{\SN}\label{sec:sn}

\SN\ is a normal type Ia supernova that exploded within or behind the disk of the host galaxy 
\object{M100} (also known as \object{NGC 4321}) in the Virgo cluster. 
Although normal, it does show some distinct properties (e.g. very high expansion velocities, 
peculiar colour evolution) that may be characteristic of type Ia supernovae arising in young 
and dusty environments \citep{2008ApJ...675..626W}.
The host galaxy recession velocity is 1571~\kms\ (at a distance of ~15.2~Mpc) 
whereas the component of the rotation velocity 
along the line of sight at the apparent SN location is about 75~\kms\
(\citealt{1995AJ....109.2444R}, \citealt{2007PASJ...59..117K}).
The latter coincides approximately with the strongly saturated \ion{Na}{i}~D and \ion{Ca}{ii}~H\&K
components and a weakly saturated CN vibrational band.
These strong components, unaffected by the supernova event, originate from an
interstellar molecular cloud (or system of clouds) within the disk of \object{M100} at some distance
in front of \SN\ \citep{2007Sci...317..924P}.
\citet{2008ApJ...675..626W} derive, via four different methods, a strong reddening of \ebv\ = 1.42 $\pm$ 0.04~mag 
(assuming a Galactic foreground extinction of 0.026~mag).

\begin{table}[t!]
\caption{Observational log for \SN\ with UVES.
The phase is given in days with respect to the $B$-band maximum light (February 20, 2006).
The UVES setting simultaneously covered three wavelength ranges
(3290-4500~\AA, 4780-5740~\AA\ and 5830-6800~\AA) with a FWHM resolution of $\sim$7~\kms.
}
\label{tb:observations}
\centering
\begin{tabular}{llll}\hline\hline
UT Date		& Phase 	& Exp. Time   & \multicolumn{1}{c}{Heliocentric Correction} \\
(dd/mm/2006)	& (days)	& (seconds)   & (\kms)		    	      \\\hline		     
18/02		& $-$2		& 4175	      & $+$14.6		    	      \\		     
06/03		& $+$14		& 8940	      & $+$7.2		    	      \\		     
22/04		& $+$61		& 15025       & $-$15.4		    	      \\		     
\hline
\end{tabular}
\end{table}

\section{Observations \& spectra}\label{sec:observations}

\SN\ was observed with the Ultraviolet and Visual Echelle Spectrograph (UVES;
\citealt{2000SPIE.4008..534D}) mounted at the European Southern Observatory (ESO) 8.2m Very Large Telescope (VLT).
Observations were carried out at four different phases with respect to 
maximum light (20 February 2006) in the $B$-band (Table~\ref{tb:observations}; see \citealt{2007Sci...317..924P}).

For UVES the 390-580 setup was used, which is unfortunate since it is notorious for not covering the best studied
DIBs at 5780 and 5797~\AA\ (even in this case where they are redshifted by about 32~\AA).
UVES data were reduced with the UVES data reduction pipeline software \citep{ballester00}.
HIRES and UVES spectra of \SN\ were taken at phase $+105$ and $+121$~days, respectively.
However, these spectra have insufficient signal-to-noise for the purpose of our analysis of weak narrow features
and are therefore not further considered in this letter.
The spectral FWHM velocity resolution is 6.6 -- 7.3~$\pm$ 0.7~\kms\ (from blue to red) and the uncertainty in the 
ThAr wavelength calibration is 0.15~\kms.
To correct for the Earth's motion a heliocentric velocity correction (Table~\ref{tb:observations}) 
has been applied to the individual exposures.

\begin{table}[th!]
\caption{Heliocentric corrected radial velocities $v$ (\kms), Doppler widths $b$ (\kms) and column densities $N$ (cm$^{-2}$) 
have been derived with VPFIT for \ion{Ca}{i}, CH$^+$ and CH transitions toward \SN. 1$\sigma$ rms-errors are given in parenthesis.
CH at phase~$+61$ is not well constrained (low S/N) but consistent with earlier phases.
Contamination in CH$^+$ (4232~\AA) at phase~$-2$ and strong noise in CH$^+$ (3957~\AA) at phase~$+14$ 
made it impossible to properly constrain fits to this lines. 
Straightforward peak velocity and strength measurements do however agree with the other two transitions.
For the CN R(0) transitions $v$ is 1645 and 1647~\kms at phase $-2$ and $+14$, respectively.
}
\label{tb:results_gastracers}
\centering
\begin{tabular}{llccr@{\,}l}
\hline\hline
Transition	&  Phase	& $v$     	& $b$	    &\multicolumn{2}{c}{ log\,$N$}  \\
		&  (days)       & (\kms)  	& (\kms)    &\multicolumn{2}{c}{(cm$^{-2}$)}\\ \hline
\ion{Ca}{i}	&  $-$2	        &   1641   (2)  & 2   (2)   & 10.8    &   (0.2)   \\
		&  $+$14        &   1643   (1)  & 1   (1)   & 10.8    &   (0.2)   \\
CH$^+$		&  $-$2	        &   1647   (1)  & 4   (2)   & 13.7    &   (0.1)   \\
		&  $+$14        &   1651   (1)  & 3   (2)   & 13.6    &   (0.1)   \\
CH		&  $-$2	        &   1645   (1)  & 4   (1)   & 13.78   &   (0.03)  \\
		&  $+$14	&   1647   (1)  & 2   (1)   & 13.8    &   (0.3)   \\
		&  $+$61        &   1648   (2)  & 2   (2)   & 13.8    &   (0.2)   \\ \hline
\end{tabular}
\end{table}

\begin{table}[th!]
\caption{Radial heliocentric velocities and equivalent widths for $\lambda\lambda$~6196 and 6283 ($1\sigma$ rms-error in parenthesis).
3$\sigma$ (=3.0~FWHM~$\sigma_{\rm noise}$) upper limits for $\lambda\lambda$~6379 and 6613.
Uncertainties in $\lambda$6283 central velocity are of the order of 10 to 20~\kms.
}
\label{tb:results_dibs}
\centering
\resizebox{\columnwidth}{!}{
\begin{tabular}{lcccccc}\hline\hline
            & \multicolumn{2}{c}{6196}     	   & \multicolumn{2}{c}{6283}       & 6379        & 6613	       
	    \\\cline{2-3}\cline{4-5}\cline{6-6}\cline{7-7}
Phase	    & $v$      	    	&  $w$		   & $v$      	   &  $w$	    &  3$\sigma$  & 3$\sigma$   \\
(days) 	    & (\kms)   	    	& (m\AA)	   & (\kms) 	   & (m\AA)	    &  (m\AA)     & (m\AA)	\\\hline
$-$2	    & 1647 (1)  	& 12 (2)   	   & $\sim$1677    & 160 (15)       &  $<$ 13	  & $<$ 31  	\\  
$+$14	    & 1650 (1)  	& 13 (2)           & $\sim$1656    & 180 (15)       &  $<$ 11	  & $<$ 29  	\\  
$+$61	    & 1649 (1)  	& 16 (3)           & $\sim$1639    & 190 (20)       &  $<$ 25	  & $<$ 55  	\\  
\hline
\end{tabular}
}
\end{table}

\section{Interstellar atoms, molecules and diffuse bands}\label{sec:results}
The \ion{Ca}{i} , CH$^+$ and CH transitions were fitted with the VPFIT code\footnote{%
VPFIT (version 9.3) has been developed by R.\,F.~Carswell, K.~Webb, M.\,J.~Irwin, and A.\,J.~Cooke. 
It is available at http://www.ast.cam.ac.uk/$\sim$rfc/vpfit.html}.
The resulting central velocities and column densities are given in Table~\ref{tb:results_gastracers} 
for \ion{Ca}{i}, CH$^+$ and CH. VPFIT uncertainties depend directly on the line parameters (width and 
strength) and the S/N. The smallest Doppler widths ($\sim$1~\kms) indicate that these lines are not, 
or only marginally, resolved.
The small shift in velocity between \ion{Ca}{i} and CH$^+$/6196~DIB is notable but
probably systematic due to 
their relative weakness and presence of noise artifacts near the 
interstellar features. 
The $\sim$2~\kms\ velocity difference between the first two epochs is probably due to a rigid shift of the UVES instrument between nights.

Template 6196, 6283, 6379 and 6613~\AA\ DIBs were constructed (via a multiple Gaussian profile fit)
from high quality ($R \sim 80\,000, S/N \sim 600$) spectra of the Galactic target $\beta^1$\,Sco
 (\object{HD144217}; \ebv\ = 0.22~mag).
The 6196 and 6283 template profiles were fitted to the observed spectra with intensity and central
velocity set as free parameters.
The 6379 and 6613~\AA\ DIBs are not detected, but upper limits on the strength (at the expected velocity) were derived. 
Central velocities and equivalent widths or 3$\sigma$ limits are reported in Table~\ref{tb:results_dibs}.
The central velocities of the 6283~\AA\ DIB are uncertain due to the width of the profile and the large uncertainty
on its rest wavelength, $\lambda_0$.

\section{Discussion \& conclusion}\label{sec:discussion}
The high quality (in particular high S/N) of the UVES data has made it possible 
to detect and analyse molecular lines and diffuse interstellar bands toward \SN\ probing
the ISM of \object{M100}.

The (dense) gas tracers \ion{Ca}{i}, CH and CH$^+$ are detected in the spectra of \SN\ 
taken at phase $-2$ and $+14$. CH has also been detected in the phase $+61$ spectrum.
Non-detections for later phases are due to reduced S/N.
Radial velocities and column densities remain invariant (within error bars) between the
different phases.
The derived CH column density is in good agreement, using the Galactic correlation (\citealt{2004A&A...414..949W}),
with that derived from CN and \ion{Na}{i} measurements, implying \ebv\ $\sim$~1.
The lower CH$^+$ abundance suggests a quiescent medium without the frequent shocks
that are normally associated with CH$^+$ production (\citealt{1989MNRAS.241..575C}, \citealt{1995A&A...300..870F}).

The equivalent widths of the detected DIBs are also invariant within the uncertainties.
The observed $\lambda$6283 is seven times weaker (adopting \ebv\ = 1.4 for \SN) 
compared to the average Galactic (MW) value, while the observed $\lambda$6196 is about five times weaker.
See Table~\ref{tb:dibs} for a comparison of different Galactic and extra-Galactic lines-of-sight.
The non-detected $\lambda\lambda$6379 and 6613 are at least a factor of fifteen under-abundant.
Since $\lambda$6196 and $\lambda$6613 are closely correlated in Galactic environments it
is striking that $\lambda$6196 is detected while the usually stronger $\lambda$6613 is not.
From the observed DIB strengths it is not possible to determine how much more under-abundant the 
6379 and 6613 DIBs are with respect to the detected 6196 and 6283 DIBs (whose strength ratio is in the observed range).
The overall weakness of DIBs toward SN2006 suggests that the general relation between DIB strength 
and reddening does not hold.

\begin{figure}
\includegraphics[bb=40 100 255 415,angle=-90,width=.95\columnwidth,clip]{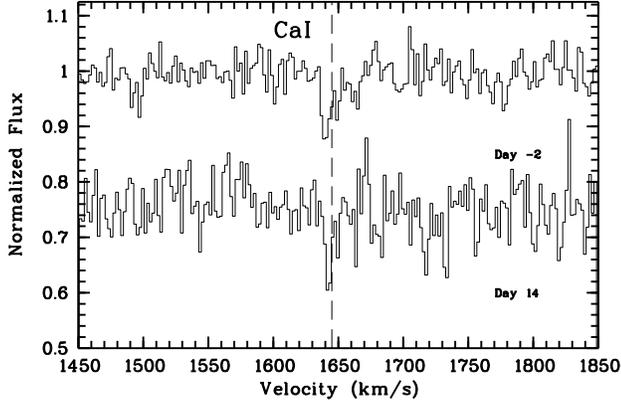}
\label{fig:CaI}
\caption{%
The \ion{Ca}{i} ($\lambda_{\rm 0}\ = 4226.73$~\AA) transitions toward \SN. 
Spectra are rebinned to 2~\kms. Vertical dashed line at $v_{\rm helio}$~=~1645~\kms.
}
\end{figure}

\begin{figure}
\includegraphics[bb=40 100 255 415,angle=-90,width=.95\columnwidth,clip]{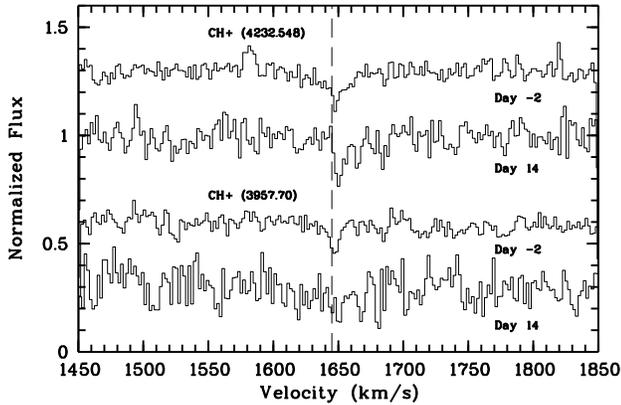}
\label{fig:CHp}
\caption{%
The CH$^+$ (top two spectra with $\lambda_{\rm 0}~=~4232.548$~\AA\
and bottom two spectra with $\lambda_{\rm 0} = 3957.70$~\AA) transitions toward \SN. 
Spectra are rebinned to 2~\kms. Vertical dashed line at $v_{\rm helio}$~=~1645~\kms.
}
\end{figure}

\begin{figure}
\includegraphics[bb=40 100 255 415,angle=-90,width=.95\columnwidth,clip]{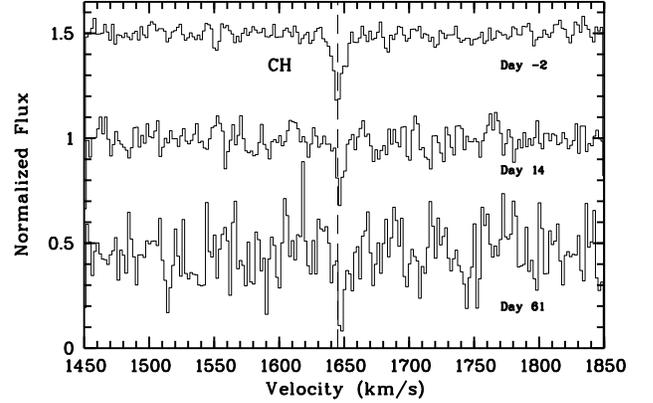}
\label{fig:CH}
\caption{%
The CH transition toward \SN\ ($\lambda_{\rm 0}~=~4300.303$~\AA).
Spectra are rebinned to 2~\kms. Vertical dashed line at $v_{\rm helio}$~=~1645~\kms.
}
\end{figure}

\begin{figure}
\includegraphics[bb=40 100 240 415,angle=-90,width=.95\columnwidth,clip]{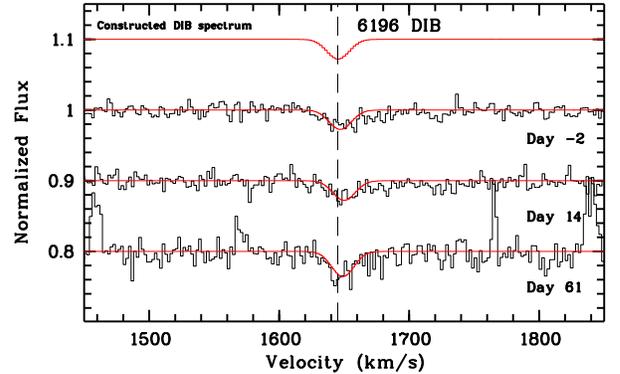}
\label{fig:6196}
\caption{%
The 6195.96~\AA\ DIB detection toward \SN. 
The reference spectrum (top) has been constructed from a Galactic 6196 band profile shifted
to the \SN\ velocity. 
Spectra are rebinned to 2~\kms. Vertical dashed line at $v_{\rm helio}$~=~1645~\kms.
}
\end{figure}

\begin{figure}
\includegraphics[bb=40 100 315 415,angle=-90,width=.95\columnwidth,clip]{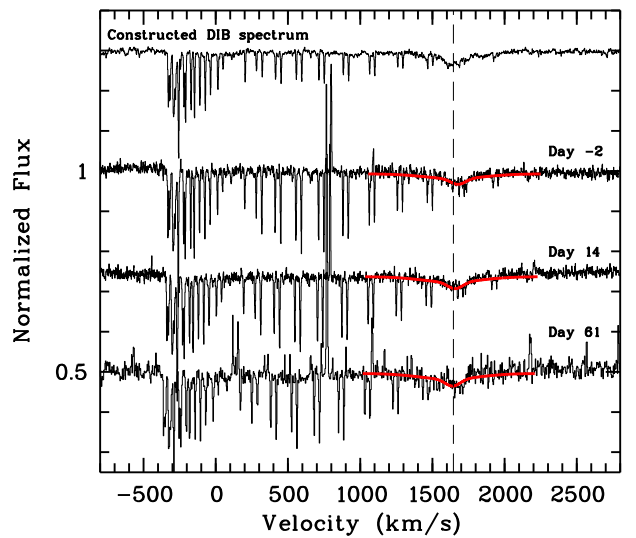}
\label{fig:6284}
\caption{%
The 6283.85~\AA\ DIB is detected toward \SN. 
The reference spectrum (top) has been constructed from the Galactic 6283 band profile 
of $\beta^1$\,Sco and a Galactic telluric (unreddened) standard spectrum. 
The spectrum at phase $-2$ is rebinned to 2~\kms\ and to 4~\kms\ at $+61$ days.
Vertical dashed line at $v_{\rm helio}$~=~1645~\kms.
}
\end{figure}

\begin{figure}[th!]
\includegraphics[bb=40 100 255 415,angle=-90,width=0.925\columnwidth,clip]{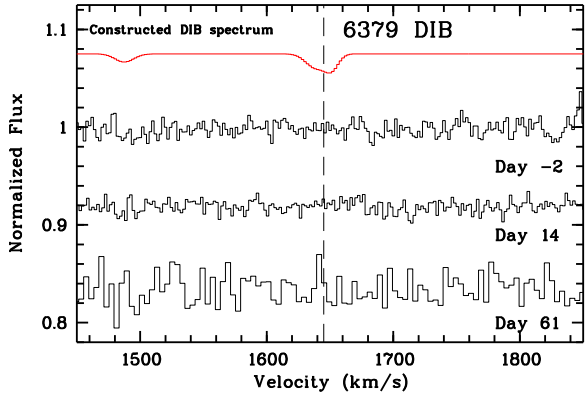}
\label{fig:6379}
\caption{%
The 6379.6~\AA\ DIB is not detected toward \SN. 
The reference spectrum (top) has been constructed from the Galactic 6379 band profile toward
$\sigma$\,Sco shifted to the \SN\ velocity.
Spectra at phase $-2$ and $+14$ are rebinned to 2~\kms, and
the spectrum at phase $+61$ to 4~\kms.
 Vertical dashed line at $v_{\rm helio}$ = 1645~\kms.}
\end{figure}

\begin{table}[t!]
\caption{Comparison of DIB strengths (in m\AA) toward \SN\ in M100, the Galaxy and LMC. 
Upper limits for $\lambda\lambda$6379 and 6613 obtained from weight averaged spectra at phase $-$2 and $+$14.
Equivalent widths for DIBs toward $\beta^1$\,Sco and $\rho$\,Oph from \citet{2005A&A...429..559S}
and for \object{Sk-69 223} from \citet{2006A&A...447..991C}.
Taurus dark cloud toward \object{HD283809} from \citet{1991MNRAS.252..234A}.}
\label{tb:dibs}
\centering
\resizebox{\columnwidth}{!}{
\begin{tabular}{lllllll}\hline\hline
DIB		& \SN\		&  MW 	   & $\beta^1$\,Sco& $\rho$\,Oph   &\object{Sk-69 223} & \object{HD283809} \\  
\ebv		&  $\sim$ 1	&  1.0     &	0.22	   & 0.32	   &  0.35	    & 1.6	 \\  \hline
6196		& 14 $\pm$ 4    & 53	   & 20 	   & 10 	   &  10	    & 46	 \\  	
6283		& 177 $\pm$ 25	& 900 	   & 390	   & 111	   &  225	    & 312	 \\     
6379		& $\la$ 8	& 88       & 14 	   & 24 	   &  55	    & --	 \\  		       
6613		& $\la$ 18	& 210	   & 40 	   & 43 	   &  19	    & --	 \\   
\hline								
\end{tabular}							
}
\end{table}

The inferred total-to-selective visual extinction ratio, $R_V$, of the dust toward \SN\ is 1.48 \citep{2008ApJ...675..626W}, 
much lower than the typical values observed for Galactic extinction: $R_V$ = 3.1 $\pm$ 1.5 \citep{2007ApJ...663..320F}.
Therefore, if we compare DIB strengths per unit visual extinction (with $A_V$ = $R_V$ $\cdot$ \ebv) the
apparent under-abundances is reduced by a factor of two.

At this point a word of caution is warranted regarding the derived reddening from photometric observations.
The presence of circumstellar light echos can affect the derived colours and subsequent determination of
\ebv\ and $R_V$ (e.g. \citealt{2005ApJ...635L..33W},
\citealt{2006MNRAS.369.1949P}), provided that dust can survive the strong radiation field generated
by the explosion.
More and more supernova studies reveal dust with very low $R_V\ \sim 1$ -- 2
(see for example \citealt{2000ApJ...539..658K}, \citealt{2006MNRAS.369.1880E,2008MNRAS.384..107E}, 
\citealt{2006ApJ...653..490W} and \citealt{2007arXiv0712.1155N}).
It has not been established yet if the observed peculiar dust extinction is directly connected 
to the supernovae events or are representative of the extra-galactic ISM in the host galaxies.

Whether the remaining under-abundance of DIBs with respect to dust extinction is due to a lower gas-to-dust
ratio (i.e. relatively less gas and more dust in the ISM), lower metallicity (i.e. not enough carbonaceous 
material to build DIBs) and/or a hard radiation field (i.e. a higher DIB destruction rate) 
is impossible to conclude without additional information on this line-of-sight (see for example \citealt{2006A&A...451..973C}).

One tantalizing possibility is that the line-of-sight toward \SN\ probes a dense interstellar cloud whose interior
is shielded from the UV radiation and shows increased grain growth. 
\citet{1991MNRAS.252..234A} found a remarkable decline in DIB strength (per unit reddening) with increasing extinction
(see Table~\ref{tb:dibs} for values for the dark cloud toward \object{HD283809}).
It is unclear if the reduced strength of DIB features in dark clouds points to a grain or gas-phase related carrier.

\begin{figure}[t!]
\includegraphics[bb=40 100 255 415,angle=-90,width=.925\columnwidth,clip]{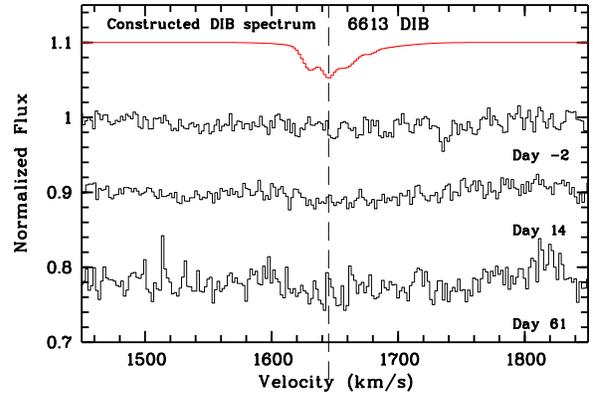}
\label{fig:6613}
\caption{%
The 6613.6~\AA\ DIB is not detected toward \SN. 
The reference spectrum (top) has been constructed from the Galactic 6613 band profile
toward $\sigma$\,Sco shifted to the \SN\ velocity. 
Spectra are rebinned to 2~\kms. Vertical dashed line at $v_{\rm helio}$~=~1645~\kms.
}
\end{figure}

Infrared/sub-mm data shows that \object{M100} is a typical spiral galaxy with a metallicity 
close to solar, a Galactic like PAH-to-dust mass fraction and a dust-to-gas mass ratio about twice that of the Galaxy 
\citep{2007ApJ...663..866D}. 
From this point of view, keeping in mind it is an overall average of the entire \object{M100} galaxy, there is twice the amount of
dust compared to gas in \object{M100} than in the Milky Way (thus also the lower reddening inferred from CH and CN).
This could have a further effect, such as a decrease, on the DIB strength versus reddening relation.

Optically bright supernovae events offer not only a unique opportunity to study supernovae themselves
and their immediate surroundings but also provide powerful background sources for the study of the diffuse to dense
interstellar medium of galaxies beyond the Local Group that is otherwise inaccessible.

\begin{acknowledgements}
This research has made use of NASA's Astrophysics Data System and the SIMBAD database, operated at CDS, Strasbourg, France.
\end{acknowledgements}

\bibliographystyle{aa}
\bibliography{/home/ncox/Papers/Bibtex/bibtex}  

\end{document}